\documentclass[11pt]{article}

\usepackage{epstopdf,float}
\usepackage{xurl}
\usepackage{multirow}
\usepackage{graphics}
\usepackage{amsmath,amsthm}
\usepackage{subfigure}
\usepackage{color, setspace, multirow}
\usepackage[T1]{fontenc}
\usepackage[utf8]{inputenc}
\usepackage{authblk} 
\usepackage{natbib} 
\usepackage{graphicx}
\usepackage{array}
\usepackage{mwe,hyperref}
\usepackage{graphbox, lineno}
\usepackage[export]{adjustbox}
\usepackage[normalem]{ulem}
 \usepackage{enumitem}\usepackage{amssymb}
\DeclareMathAlphabet{\mathbbold}{U}{bbold}{m}{n}

\newcommand{\bfU}{{\bf U}}

\newcommand{\bfW}{{\bf W}}

\newcommand{\bfX}{{\bf X}}

\newcommand{\bftheta}{\mbox{\boldmath $\theta$}}

\newcommand{\bfbeta}{\mbox{\boldmath $\beta$}}

\newcommand{\bfalpha}{\mbox{\boldmath $\alpha$}}

\newcommand{\bfgamma}{\mbox{\boldmath $\gamma$}}
\newcommand{\bflambda}{\mbox{\boldmath $\lambda$}}

\newcommand{\bfZ}{{\bf Z}}
\newcommand{\bfY}{{\bf Y}}

\newcommand{\bfA}{{\bf A}}

\newcommand\redsout{\bgroup\markoverwith{\textcolor{red}{\rule[0.5ex]{2pt}{0.4pt}}}\ULon}

\newcommand{\blind}{1}

\def\spacingset#1{\renewcommand{\baselinestretch}%
	{#1}\small\normalsize} \spacingset{1}

\begin{document}

\if1\blind
{
	\title{The Impact of a Gridded Streamflow Measure on Drought Variation in the Conterminous United States}
	\author{Rob Erhardt$^{\dagger, 1}$, Courtney Di Vittorio$^{2,1}$, Staci Hepler$^{1}$, Mostafa Shams$^{1}$, Wendy Wei$^{1}$, James Zhao$^{1}$\\
		$^{1}$Department of Statistical Sciences, Wake Forest University, U.S.\\
        $^{2}$Department of Engineering, Wake Forest University, U.S.
	}
	\maketitle
} \fi

\if0\blind
{
	\title{\bf A }
	\maketitle
} \fi

\bigskip
\begin{abstract}
Models for droughts draw on a wide range of meteorological and hydrological inputs.  Stakeholders classify droughts according to different purposes and priorities, and accordingly rely on different measures to explain and predict the onset of drought.  While many meteorological inputs are available as gridded data products, hydrological streamflow measurements are often only available as point-referenced gauge measurements.  This leads to an issue of misalignment for studies relying on both areal meteorological and point-referenced hydrological data.  Such gauge data can also have notable spatial and/or temporal missingness.  Many areas remain ungauged, and where gauges exist, equipment malfunctions cause temporal gaps.  In this study, we document the value of an existing gridded streamflow measure for the conterminous United States, one which was specifically designed to match the spatio-temporal support of publicly available meteorological and ordinal drought measurements.  We use this homogenized database to assess the relative importance of this streamflow measure in explaining US drought variability.  This assessment first requires that we address autocorrelation and variability in the variance of observed droughts, two extensions to existing statistical methodology.  After suitably controlling for both, our results show that the streamflow measure is often the most statistically important explanatory variable from among a wide set of meteorological variables.  We explore the spatial variation and some drivers of this result.  
\end{abstract}

\noindent%
\vfill

$\dagger$ Corresponding author.  Address 1834 Wake Forest Road, Winston-Salem NC, 27109, US.  Email \texttt{erhardrj@wfu.edu}

\newpage
\spacingset{1.45} 

\section{Introduction}
\subsection{Drought and Associated Environmental Variables}
The National Integrated Drought Information System (NIDIS) distinguishes between five different types of drought, which are shown in Table \ref{tab:intro}.  These definitions vary according to stakeholders’ priorities and concerns.  Classifying a location and time period as being in one of these drought types often involves a range of inputs, which could include meteorological variables such as precipitation and temperature, hydrological inputs on steamflow or water supply levels, and possibly ecological and economic inputs as well.  It is obvious that certain inputs can be used to deterministically \textit{define} certain drought types, but stakeholders often rely on a range of drought inputs to capture the complex nature of the drought for their purposes.

\begin{table}[!h]
  \centering
    \begin{tabular}{l|p{4.5in}}
    Type & Description\\
     \hline
     Meteorological  & When dry weather patterns dominate an area\\
     \hline
     Hydrological & When low water supply becomes evident in the water system.\\
     \hline
     Agricultural & When crops become affected by drought.\\
     \hline
     Socioeconomic & When the supply and demand of various commodities is affected by drought\\
     \hline
     Ecological & When natural ecosystems are affected by drought\\
    \end{tabular}%
    \caption{Drought categories and descriptions from the National Integrated Drought Information System (NIDIS, \url{https://www.drought.gov/what-is-drought/drought-basics}).}
  \label{tab:intro}%
\end{table}%

\begin{figure}[h!]
  \centering
\includegraphics[width=5in]{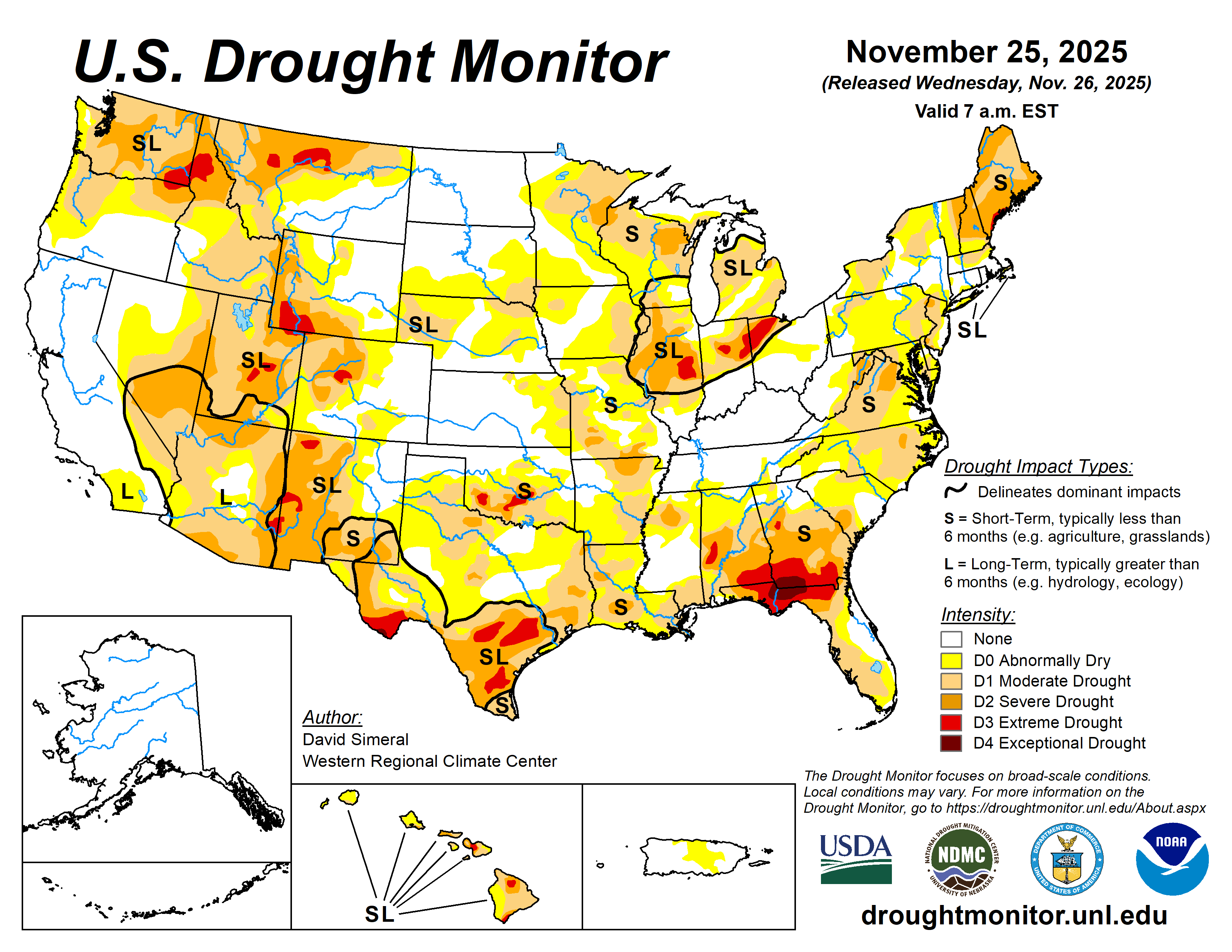}
    \caption{US Drought Monitor as of November 25, 2025.  The U.S. Drought Monitor is jointly produced by the National Drought Mitigation Center at the University of Nebraska-Lincoln, the United States Department of Agriculture, and the National Oceanic and Atmospheric Administration.  Map courtesy of NDMC.}
  \label{fig:usdmcurrent}
\end{figure}

One comprehensive measure of drought is the United States Drought Monitor (USDM, \cite{svoboda2002drought}). 
The monitor classifies regions of the U.S. into one of six ordered categories: 0, no drought; D0, abnormally dry or ``pre-drought''; and levels D1 through D4, which represent increasing levels of severity of drought.  Figure \ref{fig:usdmcurrent} shows one example of the USDM for November 25, 2025.  The USDM communicates drought severity using a categorical variable informed by several indicators, including, but not limited to: temperature, precipitation, soil moisture, snow cover, streamflow, and water levels.  A group of experts use their best judgment to determine regional drought status and confer with other state and local experts before publishing weekly drought maps.  Accordingly, the map is a blend of environmental drivers, local expertise, and drought types, all of which differ across regions.

Many inputs for drought measures, including the USDM, are meteorologic measurements, generally available through gridded data products with consistent spatial and temporal coverage.  Other inputs are hydrological, such as streamflow and water levels, and these are typically point-based measurements obtained from gauges and may contain large gaps in space or time \citep{andrews2024strategic}.  When studying the relative importance of a range of meteorological and hydrological inputs, there is an immediate spatial misalignment issue, as well as challenges with infilling missing data, particularly for hydrological measurements which rely more on in-situ gauges and equipment as opposed to remote sensing or reanalysis models. Additionally, many meteorologic and hydrologic variables used as drought indicators are often correlated with one another.  Meteorological conditions --- specifically temperature and precipitation --- are drivers of basin-scale hydrologic drought \citep{van2012process}, which in turn impacts streamflow and water levels. However, the relationship between dry and warm conditions and reduced streamflow and water levels is complex and dynamic in basins where the water system is heavily engineered and regulated \citep{tijdeman2018human}.  Water managers use streamflow to predict drought, and subsequently alter streamflow to reduce the negative socio-economic and ecologic impacts of drought.  The magnitude of this correlation can depend on the extent to which a catchment’s natural water balance is altered and managed \citep{tijdeman2018natural, haslinger2014exploring}.  Accordingly, a second challenge beyond the spatial misalignment issue is how to build models to properly account for dependencies among input measurements and local water management practices.  

Streamflow in particular is an important variable for drought assessment, and an increased spatial and temporal resolution of streamflow data is expected to enhance drought predictions \citep{brunner2021challenges}. 
NOAA's National Water Center has been developing the National Water Model to produce spatially continuous streamflow estimates in near real time, and produce medium-range (10-day) and long-range (30-day) streamflow forecasts \citep{cosgrove2024noaa}.  
We therefore expect enhanced streamflow databases to become available in the near term, and seek to understand the relationship between streamflow and drought in gauged versus ungauged locations.

Our goal is to quantify the relative statistical importance of streamflow measurements in a comprehensive statistical model for drought in the conterminous United States. Given what we have described, this requires a single homogenized database, blending drought, hydrological and meteorological measurements for the full conterminous United States, with multiple years of data to capture a wide range of observed drought outcomes.  Fortunately, \cite{erhardt2024homogenized} previously built and published such a database.  Their data combine gridded ordinal drought measurements from the USDM defined weekly for the conterminous United States, along with seven environmental measurements.  These seven measurements are: accumulated precipitation, evapotranspiration, potential evapotranspiration, soil moisture, soil surface runoff, soil temperature, and a gridded measurement of streamflow in the previous 28 days.  
Of these, the first six are drawn from the North America Land Data Assimilation System Phase 2 (NLDAS-2), an integrated observation and model reanalysis data set \citep{mitchell2004multi, xia2012continental}.  The seventh variable, the streamflow measurement over the previous 28 days, was separately constructed from in-situ stream gauge data and accordingly is expected to be less correlated with the meteorological variables which all stem from a common reanalysis model.  This measurement is based on percentiles, which have been used as a direct measure of hydrologic drought \citep{sung2014} as well as in studies on how these drought characteristics have changed over time \citep{patterson2013}.  We sought a measurement built from the 28-day average streamflow percentiles, since a monthly average would allow for the incorporation of hydrologic processes that impact drought on a longer time scale (such as snowfall melt) than the meteorological variables in the data set.

These data are freely available at \url{https://datadryad.org/stash/dataset/doi:10.5061/dryad.g1jwstqw7}.  
The full sources of raw data, preprocessing algorithms, exploratory data analyses and extensive discussion are available in \cite{erhardt2024homogenized}.  All measurements are taken from the same spatial support, which was a lattice of 0.5 degree longitude by 0.5 degree latitude covering CONUS.  We let $i=1, .., I=3246$ define the spatial location in this study.  We also let $t=1, ..., T=783$ define the unique weeks between 2007 through 2021, inclusive.  These dates were chosen to ensure complete data through all time periods at the $I$ locations, as some locations had scattered missingness in streamflow measurements pre-2007 or post-2021.  While this database already exists, no comprehensive quantitative study has utilized this full database to quantify the relative statistical value of different predictors.

\subsection{Research Questions} 
In this paper, we seek to quantify the statistical importance that a gridded hydrological measurement on streamflow has on the US Drought Monitor, relative to a set of meteorological measurements.  Specifically, we ask:
\begin{enumerate}

\item What is the relative effect of streamflow compared to all meteorological covariates in explaining the variability in the US Drought Monitor?  When answering this, how do we best ensure that our model controls for temporal dependence and variability in the variance of drought across both seasons and locations?

\item Does the relative effect of streamflow depend on the presence of in-situ gauges within grids or whether the river network is heavily managed?

\end{enumerate}  
A full study of the relative value of streamflow across the entire US requires the use of the full database, with numerous variables, hundreds of time periods, and several thousand locations.  It also requires properly controlling for observed temporal dependence as well as capturing any variability in variance of drought, across both seasons as well as locations, which further increases the dimension of the parameter space and number of variables. While \citet{erhardt2024spatio} conducted an earlier study modeling the USDM in terms of meteorological covariates published in \citet{erhardt2024homogenized}, the computational complexity of their statistical model limited its application to only a few hundred locations and time periods, and only three meteorological predictors. It could not scale to the size data we require to address research question 1.  Accordingly, in this paper we develop a different statistical model capable of fully addressing research question 1. 

The remainder of this paper is organized as follows.  Section 2 performs some exploratory analysis.  Section 3 describes the Bayesian statistical model to address research question 1, and demonstrates the best autoregressive order and model for capturing variability in the variance.  Section 4 shows extensive results answering research question 1.  Sections 5 and 6 then describe additional statistical models and methodology needed for research question 2, with results following.  We end with some overall discussion in Section 7.   

\section{Some Exploratory Data Analysis}
Our first research question requires that we capture variability in the variance of drought, both across locations but also across seasons.  This requires us to first document the empirical variability of drought across both locations and seasons, so that it can be appropriately modeled. We call the spatiotemporal drought data $Y_{i,t}$, which simply maps each ordered level of drought to the integers in $\{0, 1, 2, 3, 4, 5\}$, with 0 $\rightarrow 0$, D0 $\rightarrow 1$, D1 $\rightarrow 2$, and so forth.  We computed the empirical variance of drought $Y_{i,t}$ by grid cell and by season, with season defined by combining adjacent months DJF, MAM, JJA, and SON to stand in for winter, spring, summer, and fall.  Specifically, to compute the winter variance we fixed a grid cell $i$ and took the set of $Y_{i,t}$ for $t \in \{\text{DJF}\}$ across all years, and computed the empirical variance for that single location.  This was repeated for all locations $i=1, ..., I$, and again for the remaining three seasons.  Results are shown below in Figure \ref{fig:vars}.  There is a pattern of change in variance by season, most notably in the southern region of the US, the Midwest, and the southwest coastal region. For example, in the south, the variability in drought is moderately high in the fall, decreases in the winter, reaches a minimum in the spring before increasing again in the summer. Similarly, the variance in the soutwest coastal region is high in the fall, lower in the winter, moderately high in the spring and finally highest in the summer.  Additional plots depicting the differences in empirical variance between each of the seasons are in the appendix (Figures \ref{fig:fall}, \ref{fig:wint}, \ref{fig:spring}, \ref{fig:summ}).

\begin{figure}[h!]
  \centering
\includegraphics[width=6in]{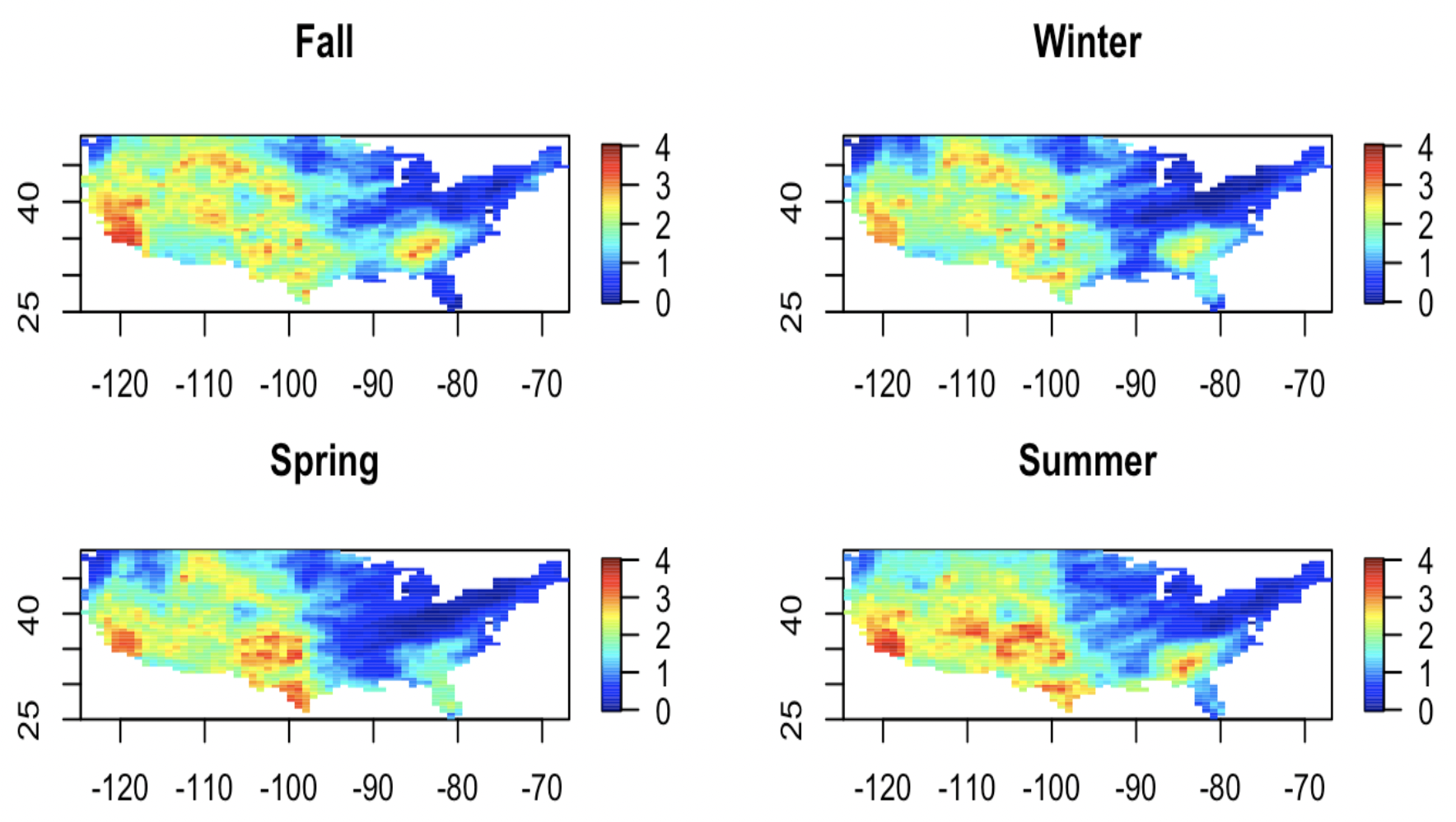}
    \caption{Empirical variance of drought $Y_{i,t}$ by season.  Observe both location-specific differences as well as overall seasonal differences.}
  \label{fig:vars}
\end{figure}

\section{Methods for Research Question 1}
\subsection{Model for Ordinal Drought}
Our model for the first research question is an extension of a spatio-temporal model first published in \cite{erhardt2024spatio}.  While we use a similar structure, we extend the model in three important ways to capture higher orders of temporal autocorrelation, achieve greater computational efficiency needed to analyze the full scale of our data, and allow for varying precision in the latent drought process with additional covariates.

We assume ordinal drought $Y_{i,t} \in \{0,1,...,J\}$, which takes one of $J+1$ ordered levels, for locations $i=1, ..., I$ and time periods $t=1, ..., T$.  We assume a latent, continuous process $Z_{i,t}$ such that $$Y_{i,t} = \sum_{j=0}^J j \cdot \textrm{I}(\alpha_j < Z_{i,t} \leq \alpha_{j+1}),$$ 
where $\bfalpha = (-\infty, 0, 1, 2, 3, 4, \infty)$.  Observe that the first category of this ordinal response is zero instead of one.  Previous research has shown that placing a prior distribution on $\bfalpha$ and modeling this parameter as part of the Bayesian model leads to computational challenges \citep{schliep2015data}, and that this problem grows quickly with the number of locations $I$, time periods $T$, and ordinal levels $J+1$, all of which are large in our setting.  \citet{erhardt2024spatio} treated this as a parameter to be fit in MCMC, but had to substantially restrict the number of locations and time periods to make the model computationally feasible.  Further, the parameter $\bfalpha$ is the only common, shared parameter across locations.  By fixing $\bfalpha$ in our model here, we can run the MCMC for each location's model independently and in parallel, substantially reducing the computational cost of sampling from the posterior and therefore making it possible to fit our model across all locations in the conterminuous United States for many more time periods. 

The latent continuous process $Z_{i,t}$ is related to external covariates in $\bfX_{i,t}$ through parameters as
\begin{equation} 
Z_{i,t} = \begin{cases} 
 \bfX'_{i,t} \bfbeta_{i} + \epsilon_{i,t} & \text{ for } t=1, \\
  \bfX'_{i,t} \bfbeta_{i} + \rho_{i,1} \left(Z_{i, t-1} -\bfX'_{i,t-1} \bfbeta_{i} \right)+ \epsilon_{i,t} & \text{ for } t=2, \\
  ... & \text{ ... }\\
 \bfX'_{i,t} \bfbeta_{i} + \sum_{k=1}^K \rho_{i,k} \left(Z_{i, t-k} -\bfX'_{i,t-k} \bfbeta_{i} \right) + \epsilon_{i,t}
 & \text{ for } t > K,
\end{cases}
\label{eq:mu}
\end{equation}
where $\bfX_{i,t}$ is a $(P+1)$-dimensional vector of covariates, and $\bfbeta_{i} = (\beta_{0i}, \beta_{1i}, ..., \beta_{Pi})'$ is a location-specific $(P+1)$-dimensional parameter vector.  The errors were assumed normally distributed and independent as $\epsilon_{i,t} \overset{ind}{\sim}\mathcal{N}(0, \tau^{-2}_{i})$ with location-specific precisions $\tau^2_{i}$. The location-specific autoregressive parameters $\rho_{i,k}$ cover up through order $K$.  These spatially-varying autocorrelation parameters $\rho_{i,k}$ are transformed as $\gamma_{i,k} \equiv \text{logit}(\rho_{i,k})$ which brings the parameter space of autocorrelations to the whole of the real line $\mathbb{R}$.  We will utilize this reparameterization for the K-dimensional vector $\bfgamma_i = (\gamma_{i,1}, ..., \gamma_{i,K})'$ in the MCMC algorithm.  The spatially-varying autocorrelation parameters $\rho_{i,k}$ were given uniform(0,1) priors, which restricts them to positive autocorrelations only, meaning previous periods whose drought level was unusually high given the covariates would be more likely to be unusually high again in the next time step.  Unlike in \citet{erhardt2024spatio} which was restricted to an AR(1) model, here we consider up to AR(K).

\begin{figure}[h]
    \centering
\includegraphics[width=0.99\linewidth]{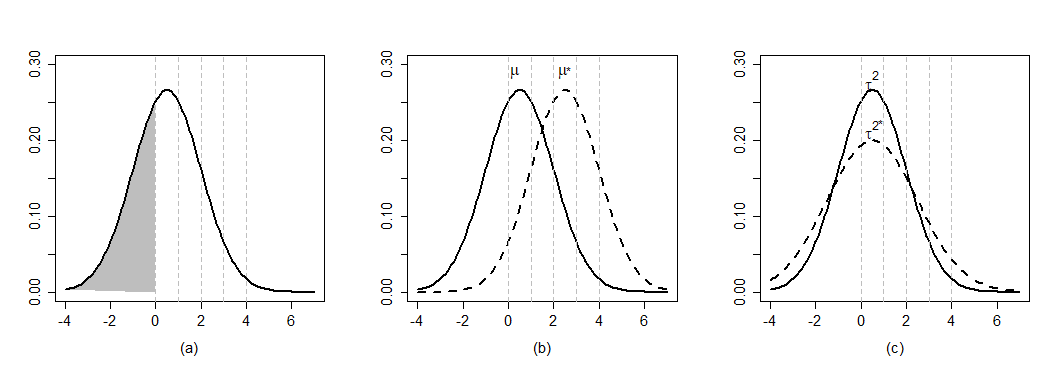}
    \caption{Illustration of the normal density of the latent variable $Z_{i,t}$ and fixed cutoff $\bfalpha = (-\infty, 0, 1, 2, 3, 4, \infty)$ which yields six ordered levels for the ordinal response $Y_{i,t}$.  Panel (a): Shaded area shows the probability $P(Y_{i,t} = 0)$ for an arbitrary distribution of $Z_{i,t} \sim \mathcal{N}(\mu, \tau^2)$.  Panel (b): A shift in the mean from $\mu$ to $\mu^*$ yields changes in the ordinal probabilities $P(Y_{i,t} = j), \ j=\in \{0, 1, 2, 3, 4, 5\}$ with more mass at higher drought levels.  Panel (c): A change in the precisions from $\tau^2$ to $\tau^{2*}$ can similarly shift probability mass to or from the extremes.  Taken together, fixed thresholds $\bfalpha$ can produce a wide range of ordinal probabilities, so long as the mean and variance are well parameterized in terms of covariates.}
    \label{fig:latentZ}
\end{figure}

Our choice to fix $\bfalpha$ permits greater computational efficiency but restricts the flexibility of the model, and in particular how the latent Gaussian $Z_{i,t}$ can place different probabilities in each of the six categories.  To enhance model flexibility as well as to capture documented variability in the variance of drought by season, we allow for variability in the precision/variance of the latent process $Z$.  We do so by bringing in additional covariates as
\begin{equation}
\log \left(\tau_{i,t}^2\right) = \bfW'_{i,t} \bflambda_i,
\label{eq:tau}
\end{equation}
where $\bfW_{i,t}$ is an (L+1)-dimensional vector of covariates for the precision, and $\bflambda_i$ is an (L+1)-dimensional site-specific parameter vector for the precision model.  The aim is to increase model flexibility by shifting probabilities of ordinal categories when the cutoffs $\bfalpha$ are themselves fixed.  Figure \ref{fig:latentZ} shows an example of the latent variable $Z_{i,t}$, how it relates to fixed cutoffs $\bfalpha$, and how we achieve model flexibility through specifications to the mean $\mu$ and precision $\tau^2$ of the latent process.  The left panel shows how this structure yields ordinal probabilities $P(Y_{i,t} = j), \ j=0, ..., J$.  The center and right panels demonstrate the model flexibility through $\mu$ and $\tau^2$ which permits varying specifications of probabilities for $Y_{i,t}$ despite the fixed threshold $\bfalpha$.  In particular, the flexibility shown in the right panel is an extension over the model in \citet{erhardt2024spatio}.

We write the Bayesian Hierarchical Model for clarity.  We define $\bfY_{1:T}$, $\bfZ_{1:T}$, $\bfX_{1:T}$ and $\bfW_{1:T}$ as observed ordinal data, latent variables, and observed covariates for the mean and observed covariates for the precision, all for all locations and time periods $1:T$.  We define the site-specific parameters as $\bftheta_{Z,i} = (\bfbeta_i, \bfgamma_i, \bflambda_i), i=1, ..., I$, and we define the full collection of all site-specific parameters as $\bftheta_Z = (\bftheta_{Z, i}, i=1, ..., I)$.  For clarity: the site-specific parameter $\bftheta_{Z,i}$ has $P+K+L+2$ parameters, covering the $P+1$ covariates for the mean, the $K$ autoregressive terms, and the $L+1$ covariates for the precision; the full collection of all parameters $\bftheta_Z$ has $I \times (P+K+L+2)$ parameters.  
Our full Bayesian Hierarchical Model is:
\begin{align*}
        \text{Data Model: } & [\bfY_{1:T} \mid \bfZ_{1:T}]\\
        \text{Process Model: } & [\bfZ_{1:T} \mid \bftheta_Z, \bfX_{1:T}, \bfW_{1:T}] \\
        \text{Prior Model: } & [\bftheta_Z]
    \end{align*}
with posterior distribution {
    \begin{align*}
        [\bfZ_{1:T}, \bftheta_Z \mid \bfY_{1:T}, \bfX_{1:T}, \bfW_{1:T} ] & \propto [\bfY_{1:T} \mid \bfZ_{1:T}] [\bfZ_{1:T} \mid \bftheta_Z, \bfX_{1:T}, \bfW_{1:T}][\bftheta_Z].
    \end{align*}}
    
\subsection{Selection of the Autoregressive Order}
The model for the mean shown in equation \eqref{eq:mu} is an autoregressive model of order $K$, and so we had to determine the optimal order $K$.  To address this, we fit a preliminary model without any consideration of variability in variance over time.  That is, we fixed $\tau_{i,t}^2 = \tau_{i}^2$ where $\tau_i^2$ is a site-specific constant that does not depend on any covariates.  We also set the covariates in $\bfX_{i,t}$ to include the seven covariates specifically mentioned in section 1: accumulated precipitation, evapotranspiration, potential evapotranspiration, soil moisture, soil surface runoff, streamflow and soil temperature.  For the precisions $\tau_{i}^2$ we specified independent Gamma(0.01, 0.01) distributions, which are weakly informative and also permits Gibbs updates through conjugacy with a normally distributed $Z$.  For each covariate parameter $\beta_{i,p}$ we specified independent and weakly informative normal conjugate prior distributions with means 0 and precisions 0.04.  For the autocorrelation parameters $\rho_{i,k}$, $k=1, 2, ..., K$, we specified independent Uniform(0,1) distributions, which are equivalent to $\gamma_{i,k} \equiv \text{logit}(\rho_{i,k})$ following independent logistic distributions with mean zero and scale one.  Observe that all prior distributions are independent, and all yield a parameter space of $\mathbb{R}$ (after transformation for $\rho_{i,k}$), which means there are no restricted regions in the parameter space.

We ran a standard Metropolis-within-Gibbs algorithm using NIMBLE \citep{de2017programming}.  NIMBLE which allows users to build hierarchical models using the BUGS language but compiles and runs these models in the much faster C++ language.  We ran one MCMC chain for each location, with a length of 50,000, using the first 40,000 as a burn-in and thinning the remainder by 5.  Chains for all locations were run in parallel, as the model has independence across locations.  Computations were performed using the Wake Forest University (WFU) High Performance Computing Facility \citep{WakeHPC}.  Posterior means and standard deviations for a representative grid cell are shown in the Appendix in Figure \ref{fig:AR(k)}.

\begin{figure}[H]
  \centering
    \includegraphics[width=5in]{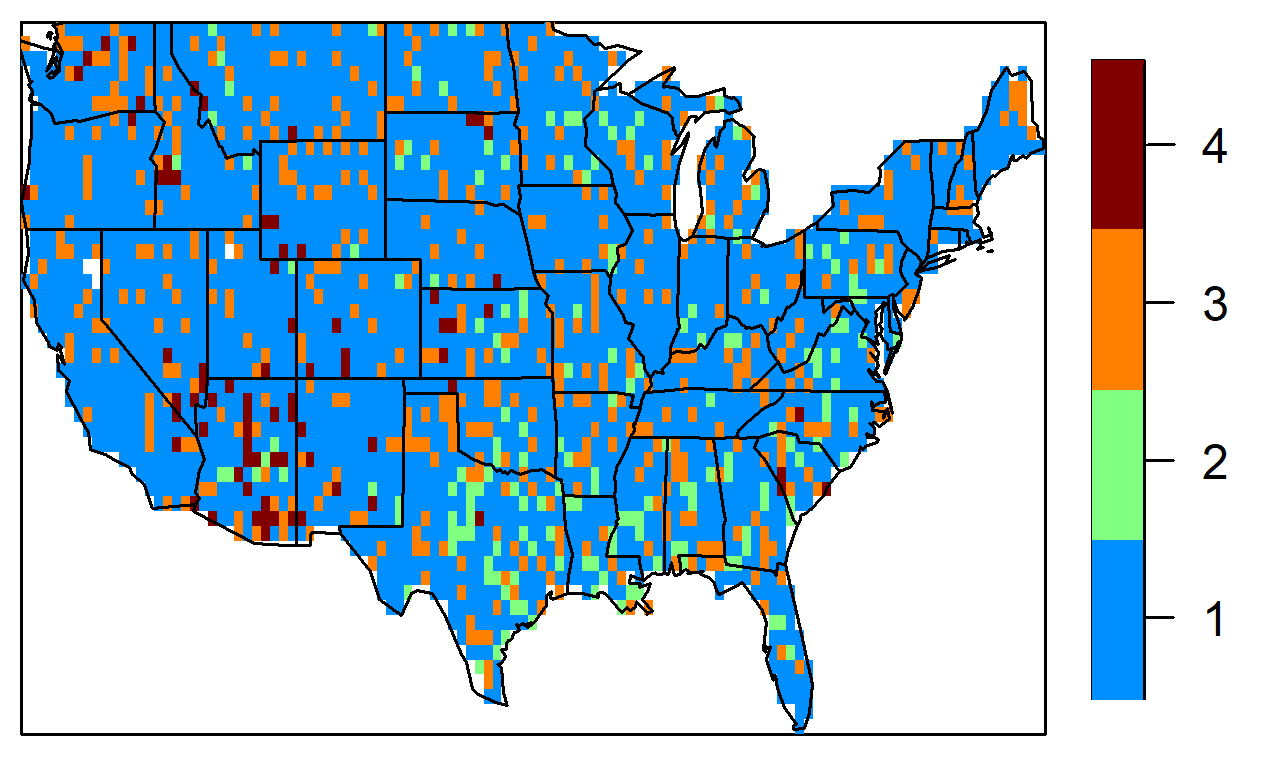}
    \caption{This plot visualizes model selection result by grid cell.  We computed WAIC for each grid cell separately, and displayed the model selected based on WAIC using different colors: blue is AR(1), green is AR(2), orange is AR(3), and red is AR(4). The vast majority of locations show the AR(1) model as best, with exceptions to this rule showing no discernible spatial pattern.}
  \label{fig:AR(p)}
\end{figure}

We considered four models, AR(1) through AR(4), and computed the Widely Applicable Information Criterion (WAIC) for each \citep{watanabe2010asymptotic}. WAIC is a Bayesian model selection criterion that estimates the out-of-sample predictive accuracy of a model while accounting for uncertainty.  We prefer the model with the lowest WAIC value.  We fit each of the four AR models and computed the WAIC by summing across all $I$ grid cells.  Results are presented in Table \ref{tab:1}.  We see that the AR(1) model achieves the lowest WAIC, and hence the best fit according to this criterion.  We also computed the WAIC for each model fit at each distinctive grid cell, $i=1, ..., I$, and selected the best model by WAIC \textit{by grid cell}.  Results are shown in Figure \ref{fig:AR(p)}.  We see that the AR(1) model is chosen for the vast majority of locations, with little spatial pattern evident from exceptions.  We therefore selected the AR(1) model for all locations in our model fitting.

\begin{table}
  \centering
    \begin{tabular}{lc}
     Model & WAIC summed over all locations\\
     \hline
     AR(1) & 1,423,467\\
     AR(2) & 1,482,984\\
     AR(3) & 1,439,426\\
     AR(4) & 1,606,478\\
    \end{tabular}%
    \caption{The total WAIC score for each of the four autorgressive order models, summed across all locations.  AR1 achieves the lowest WAIC and therefore the best fit according to this criterion.}
  \label{tab:1}%
\end{table}%

\subsection{Selection of the Varying Precision Model}
Next, we explored two parameterizations of the log-precision term $\log(\tau^2_{i,t})$ shown in equation \eqref{eq:tau}, both with the intention to produce a cyclical model with a period of one year.  In the first model, we used one sine and one cosine function which allowed the precision to change smoothly in a cyclical pattern across the year, 
\begin{equation}
\log(\tau_{i,t}^2) = \lambda_{i,0} + \lambda_{i,1} \sin \left(\dfrac{2\pi t}{52.17}\right) + \lambda_{i,2} \cos \left(\dfrac{2\pi t}{52.17}\right),
\label{eq:lambda1}
\end{equation}
where $t=1, 2, ..., T$ indexed the week and the arguments in the $\sin(\cdot)$ and $\cos(\cdot)$ functions yield a period of exactly one year.  This model had $L+1 = 3$ parameters for the precision, and necessarily achieved its maximum and minimum in opposite seasons (e.g. summer-winter, or spring-fall).  We also explored a second model for the precision with one additional parameter and more model flexibility.  This was simply a piecewise constant function across seasons, defined as
\begin{equation}
\log(\tau_{i,t}^2) = \lambda_{i,0} + \lambda_{i,1} \mathbbold{1}(t \in \text{MAM}) + \lambda_{i,2} \mathbbold{1}(t \in \text{JJA}) + \lambda_{i,3} \mathbbold{1}(t \in \text{SON}),
\label{eq:lambda2}
\end{equation}
where $\mathbbold{1}(\cdot)$ is the indicator function that takes value 1 when the argument holds, and 0 otherwise.  Notably, this second specification is not symmetric within a year and can achieve maximum and minimum variances in any combination of distinct seasons; however, it also has one additional parameter.

To select among these two models, we first fit a simplified model with $\bfbeta_i = \beta_{i,0}$ and no other covariates for the mean, and no autoregressive terms.  We defined prior distributions for $\beta_{i,0}$ and each element of $\bflambda$ as independent normal distributions with means 0 and precisions 0.04.  We sampled from the posterior using MCMC, again with the NIMBLE package \citep{de2017programming} and the DEAC cluster \citep{WakeHPC}.  Specifically, we ran a single chain for 80,000 samples, using the first 60,000 draws as a burn-in and thinning the remainder by five.  The MCMC used Metropolis-Hastings sampling to update $\bflambda_i$ and $Z_{i,t}$, while Gibbs sampling was assigned to update $\beta_{i,0}$. The run time for a single grid cell was about 6 minutes for both of the precision models when running on a single node. Note that both models shown in equations (3) and (4) were fit without any covariates $\bfX_{i,t}$, to maximize comparability with the empirical data and maintain consistency with one another.

We assessed precision model fit by utilizing a matching algorithm between empirical variances (shown in Figure \ref{fig:vars}) and modeled variances, computed as the inverse of the exponentiated modeled log-precisions.  Specifically, for each grid cell $i$, we checked if the empirical and modeled seasons for maximum variances matched, and again for minimum variances.  This process was repeated for every grid cell, for both the sin/cos model in equation \eqref{eq:lambda1} and the indicator model in equation \eqref{eq:lambda2}.  The sin/cos model was found to correctly determine the season at which the maximum empirical variance occurred for 73.937\% of the grid cells and the season at which the minimum empirical variance occurred for 80.592\% of the grid cells.  The indicator model was found to correctly determine the season at which the maximum empirical variance occurred for 86.876\% of the grid cells and the season at which the minimum empirical variance occurred for 90.049\% of the grid cells.  Based on this accuracy and computational time, we selected the indicator model shown in equation (4) for the precision.

\subsection{Overall Model Fit for Streamflow Impacts}
The previous two subsections confirmed two specific choices for the model desribed in subsection 3.1.  For clarity, we rewrite our final statistical model used to answer our first research question here.  We let

\begin{equation} 
Z_{i,t} = \begin{cases} 
 \bfX'_{i,t} \bfbeta_{i} + \epsilon_{i,t} & \text{ for } t=1 \\
 \bfX'_{i,t} \bfbeta_{i} + \rho_{i} \left(Z_{i, t-1} -\bfX'_{i,t-1} \bfbeta_{i} \right) + \epsilon_{i,t}
 & \text{ for } t > 1,
\end{cases}
\label{eq:main1}
\end{equation}
where the errors are normally distributed and independent as $\epsilon_{i,t} \overset{ind}{\sim}\mathcal{N}(0, \tau^2_{i,t})$ with location-and-time-specific precisions
\begin{equation}
\log(\tau_{i,t}^2) = \gamma_{i,0} + \gamma_{i,1} \mathbbold{1}(t \in \text{MAM}) + \gamma_{i,2} \mathbbold{1}(t \in \text{JJA}) + \gamma_{i,3} \mathbbold{1}(t \in \text{SON}).
\label{eq:main2}
\end{equation}
Using all seven covariates in $\bfX$, we fit this model to the $I = 3246$ locations for the full time period $t=1, ..., T$, independently and in parallel.  Parallelization allowed massive computational gains, and the full set of $I=3246$ models could be run in a few hours on a research computing cluster.  To assess the convergence of the MCMC sampling for our Bayesian hierarchical model, we performed a spot check convergence of key parameters for a representative sample of grid cells. Convergence was evaluated visually using trace plots and density plots (not shown).

\section{Results for Research Question 1}

\begin{figure}
  \centering
    \includegraphics[width=6in]{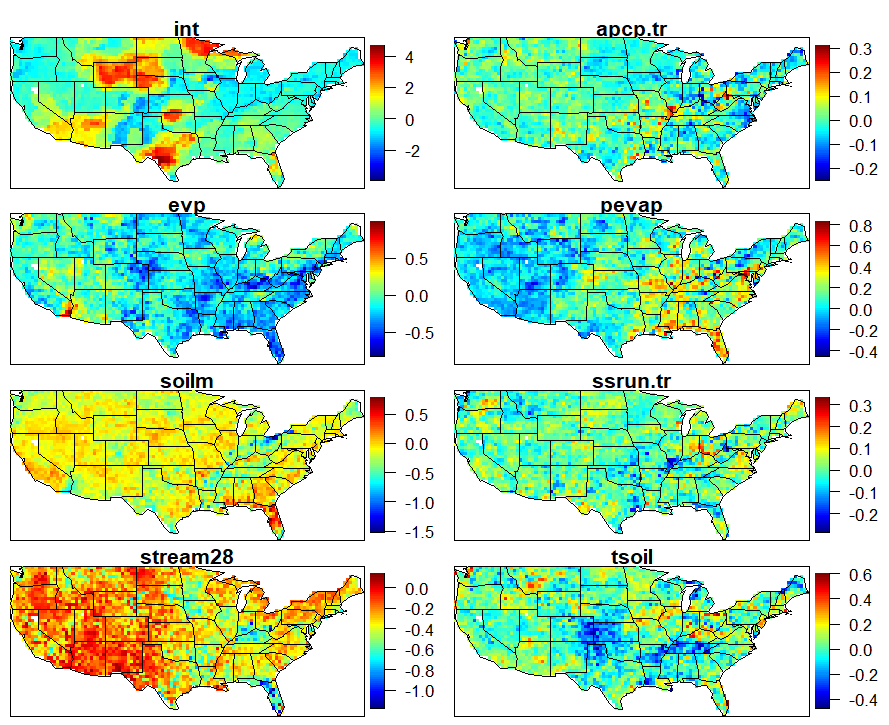}
    \caption{Posterior mean of the covariate effects in $\bfbeta$ from the model shown in equations \eqref{eq:main1} and \eqref{eq:main2}.}
  \label{fig:postmeans}
\end{figure}

\begin{figure}
  \centering
    \includegraphics[width=6in]{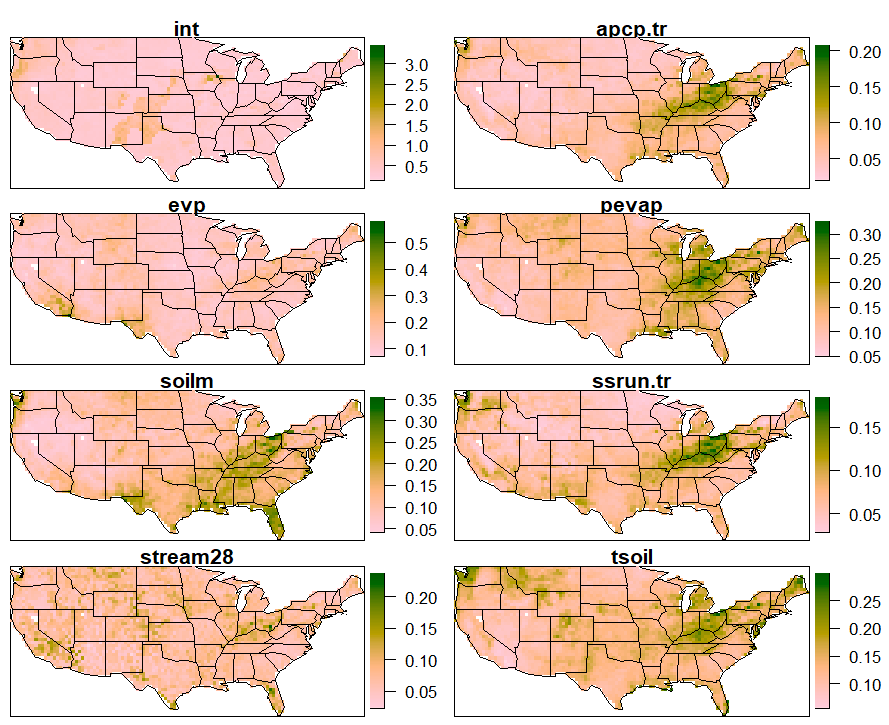}
    \caption{Posterior standard deviation of the covariate effects in $\bfbeta$ from the model shown in equations \eqref{eq:main1} and \eqref{eq:main2}.}
  \label{fig:postsds}
\end{figure}

\begin{figure}
  \centering
    \includegraphics[width=6in]{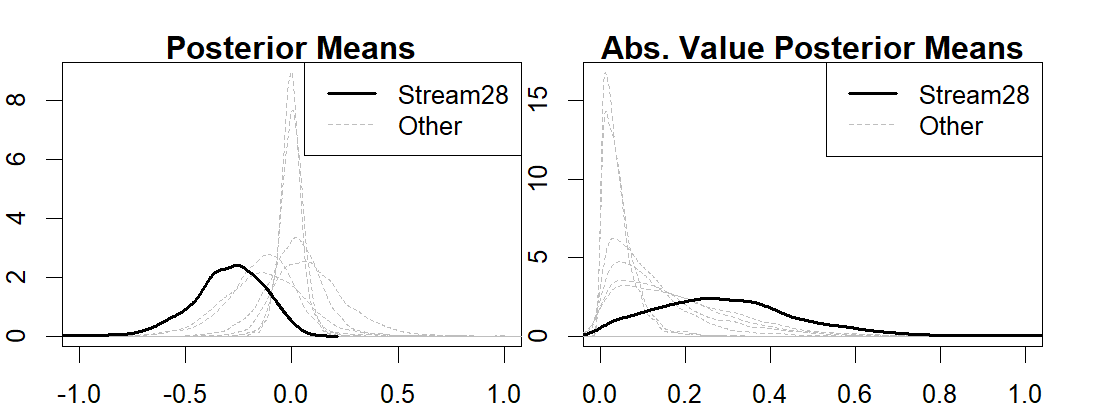}
    \caption{Smoothed kernel density estimates of the posterior means of each covariate effect, taken over the $I=3246$ distinct locations.  Left panel shows results for $\bar{\beta}_{i,p}$.  Right panel shows results for $|\bar{\beta}_{i,p}|$ for ease of interpretation.  Streamflow tends to have the largest magnitude effect among the seven covariates.}
  \label{fig:densities}
\end{figure}

Figure \ref{fig:postmeans} shows posterior means for the intercept (labelled  \texttt{int} in the figure headings) and seven covariates across space, with standard deviations shown in Figure \ref{fig:postsds}.  We abbreviate variable names as: \texttt{apcp.tr} for transformed accumulated precipitation, \texttt{evp} for evapotranspiration, \texttt{pevap} as potential evapotranspiration, \texttt{soilm} for soil moisture, \texttt{ssrun.tr} for transformed soil surface runoff, \texttt{streamflow} for the streamflow in the past 28 days, and \texttt{tsoil} for the soil temperature.  Due to the strong right skew in both accumulated precipitation and soil surface runoff, these were both transformed as described in \citet{erhardt2024homogenized}.  Recall that all covariates in $\bfX$ were standardized to all have mean zero and variance one, which makes the parameters for the seven covariates comparable to one another and on the same unitless scale.  To aid in visualizing the relative importance of $\beta_{streamflow}$, we made smooth kernel density estimates of each of the $I=3246$ posterior means, by covariate type, and we display these density estimates in Figure \ref{fig:densities}.  The left panel shows distributions for the posterior means $\bar{\beta}_{i,p}, i=1, ..., I$ and $p$ indexes the covariate, and the right panel distributions for $|\bar{\beta}_{i,p}|, i=1, ..., I$, which makes it easier to assess absolute impact regardless of sign.  The heavy black line is for $\beta_{streamflow}$, while light gray lines show distributions of the other parameters.  It is evident that $\beta_{streamflow}$ tends to have a larger magnitude effect (e.g. the distribution of posterior means is further from zero) for many locations.  The average posterior means and absolute posterior means are shown in Table \ref{tab:postmeans3}, confirming that \texttt{streamflow} is the most impactful covariate among the set.  

\begin{table}[ht]
\centering
\begin{tabular}{rrrrrrrr}
  \hline
 & apcp.tr & evp & pevap & soilm & ssrun.tr & streamflow & tsoil \\ 
  \hline
Avg. Posterior Means & -0.003 & -0.132 & 0.093 & -0.148 & 0.000 & -0.295 & 0.009 \\ 
Avg. |Posterior Means| & 0.040 & 0.186 & 0.143 & 0.176 & 0.045 & 0.297 & 0.104 \\ 
   \hline
\end{tabular}

\caption{The top row shows the average of the posterior means for each of the seven covariates, taken over all grid cells, and matches the means of each distribution shown in the left panel of Figure \ref{fig:densities}.  The bottom row shows the average of the absolute values of posterior means for all covariates, again taken over all grid cells, and matches the means of each distribution shown in the right panel of Figure \ref{fig:densities}.  We abbreviate variable names as: apcp.tr for transformed accumulated precipitation, evp for evapotranspiration, pevap as potential evapotranspiration, soilm for soil moisture, ssrun.tr for transformed soil surface runoff, streamflow for the streamflow in the past 28 days, and tsoil for the soil temperature.}
\label{tab:postmeans3}
\end{table}

\begin{figure}
  \centering
    \includegraphics[width=6in]{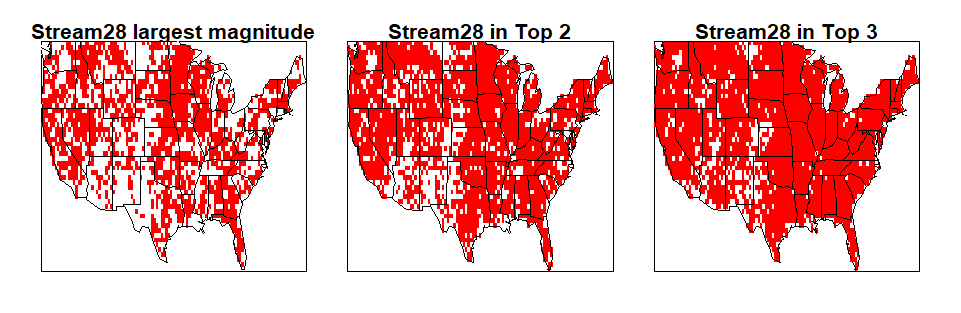}
    \caption{The locations where the absolute value of the posterior mean of streamflow is: (left) the largest; (middle) among the top two largest; and (right) among the top three largest, all from the set of seven covariates.}
  \label{fig:magnitude}
\end{figure}

Averaging across all locations, the average of posterior mean for $\beta_{streamflow}$ is -0.295.  The interpretation is that, on average, a one standard deviation change in \texttt{streamflow} results in a change of around 0.3 drought level, as the cutoffs $\bfalpha$ are spaced by 1, and so a one standard deviation increase in \texttt{streamflow} while holding all else constant would shift the mean of the latent $Z$ distribution down by around 0.3.  Figure \ref{fig:magnitude} shows which locations have $|\bar{\beta}_{i,streamflow}|$ as the single largest magnitude parameter, as one of the top two, or as one of the top three parameters from among the set of seven covariates.  We can easily see that $\beta_{streamflow}$ is often the largest magnitude parameter, and almost always in the top two or three.  

We conclude by answering our first research question.  The effect of streamflow compared to all meteorological covariates is that streamflow is the most pronounced, often with the highest magnitude parameter estimate.  Streamflow has an average effect of changing drought levels by around 1/3 of a level for every one standard deviation change in streamflow.  This result controls for autocorrelation and varying variance in the model, and holds across the conterminuous United States.  In the next section we turn to additional models needed to address our second research question --- does the relative effect of streamflow depend on the presence of in-situ gauges within grids or whether the river network is heavily managed?

\newpage
\section{Methods for Research Question 2}

\subsection{Post-Hoc Statistical Models of Streamflow}
In this next section we consider our second research question --- does the relative effect of streamflow depend on the presence of in-situ gauges within grids or whether the river network is heavily managed?  This requires that we introduce three additional binary indicators for each location.  These are: 
\begin{itemize}
\item Does the grid contain a major river within its extents?  
\item Does the grid have at least one stream gauge within its limits?  
\item Among those locations with at least one stream gauge, is the location considered ``managed'' by the USGS?  
\end{itemize}
Expanded definitions and sources of data for these measurements are described in Table \ref{tab:indicators}. 

\begin{table}[!h]
  \centering
    \begin{tabular}{l|p{5in}}
     \hline
     Variable & Definition and Source\\
     \hline
     IRiver & Does the grid contain a major river within its extents (1=yes)?  National Weather Service Rivers of the US database \url{https://www.weather.gov/gis/Rivers}\\
     \hline
     IGauge & Did the grid cell have at least one stream gauge within its limits (1=yes), or did it have no gauges and its value was drawn from an average over the smallest watershed that contained a gauge?\\
     \hline
     IManaged & Among those locations with at least one gauge (see above), does the location represent a place whose water is impacted by human management activities (1=yes)?  USGS Hydro-Climatic Data Network (NCDN 2009, \url{https://permanent.fdlp.gov/gpo22352/fs2012-3047.pdf}).\\
    \end{tabular}%
    \caption{Three binary indicator variables classifying locations for the post-hoc streamflow analysis}
  \label{tab:indicators}%
\end{table}%

To address our second research question, we fit two post-hoc spatial models to model the posterior means of $\beta_{streamflow}$ in terms of these additional binary indicators.  These indicators were used as covariates in models for the posterior means of $\bfbeta_{streamflow}$ as follows:
\begin{equation}
\bar{\beta}_{i, streamflow} \sim \mathcal{N}\left(\delta_0 + \delta_1 \mathbbold{1}(i \in \text{River}) + U_i, \tau_{\beta}^2\right)
\label{eq:post1}
\end{equation}
and
\begin{equation}
\bar{\beta}_{i, streamflow} \sim \mathcal{N}\left(\delta_0 + \delta_2  \mathbbold{1}(i \in \text{Gauged}) + \delta_3  \mathbbold{1}(i \in \text{Managed}) \mathbbold{1}(i \in \text{Gauged}) + U_i, \tau_{\beta}^2\right).
\label{eq:post2}
\end{equation}
The parameters $\delta_1$, $\delta_2$, and $\delta_3$ capture the change in effects of streamflow based on the classification of the location on river network, being gauged, or being both gauged and managed (there are no ungauged, managed locations, hence the lack of a term with the indicator for managed only in equation \eqref{eq:post2}).  Given the obvious spatial dependence across streamflow parameter estimates as shown in Figure \ref{fig:postmeans}, we include a spatial random effect in the mean.  The spatial random effect $U_i$ was given an intrinsically conditionally autoregressive (ICAR) prior.  Specifically, we defined an $I \times I$ adjacency matrix $\bfA$ where $a_{i,j} = 1$ if locations $i$ and $j$ are neighbors, and $a_{i,j} = 0$ otherwise.  This requires a binary definition of ``neighbor'', which for our areal data is simply whether or not two locations share a boundary.  Under this definition, a grid cell could have as many as 8 neighbors.  We let $a_{i+}$ be the total number of neighbors for location $i=1, ..., I$.  The ICAR prior is defined through conditional distributions as 
$$U_i \mid \bfU_{-i} \sim \mathcal{N} \left(\sum_j \frac{a_{ij}}{a_{i+}} U_j, \frac{\sigma^2_{U}}{a_{i+}} \right),$$
where $U_i \mid \bfU_{-i}$ refers to the process at location $i$ conditional on all other locations, and $\sigma^2_{U}$ is a common variance parameter.  Observe that the ICAR conditional distributions state that $U_i$ is normally distributed with a mean equal to the average of the process at the neighboring locations. 
The result is a tendency to enforce spatial smoothness in the parameter $\bfU$, allowing observations at neighboring locations to inform estimates of parameters at location $i$.  

We fit these two post-hoc models in a Bayesian framework.  The indicator variable parameters $\delta_j, j=0, 1, 2, 3$ were given diffuse independent mean-zero normal priors with standard deviations of 1000.  The precision parameters $\tau^2_\beta$ and $\tau^2_U \equiv \sigma^{-2}_U$ were given Gamma(0.01, 0.01) priors.  We again fit these two models in NIMBLE, on a single node, running a chain of length 50,000 with a burn-in of 25,000 and thinning by 5.  Results are shown in the next section.

\section{Results for Research Question 2}

Table \ref{tab:post1} shows posterior means and standard deviations for the parameter estimates from the model in equation \eqref{eq:post1}.  We see that if a location is on a major river network, the effect of streamflow changed by 0.027, with a 95\% credible interval of (0.017, 0.037).  This amounts to around 8.5\% \textit{less} of an effect of streamflow on drought status when the location is along a river network.  Table \ref{tab:post2} shows posterior means and standard deviations from the model in equation \eqref{eq:post2}, which allows for the investigation of the impact of gauged vs. ungauged locations as well as management within gauged locations.  We see that gauged locations moderate the effect of streamflow by 0.062, with a 95\% credible interval (0.044, 0.080).  This suggests that the overall effect of streamflow is not an artifact of its construction with a mixture of both observed gauged data as well as averages taken over the smallest available watersheds for ungauged grids.  Among only gauged grids, management has a nearly-negligible impact of 0.014, with a 95\% credible interval of (0.002, 0.026).

\begin{table}[ht]
\centering
\begin{tabular}{rrrrr}
  \hline
 & $\delta_0$ & $\delta_{River}$ & $\tau^2_\beta$ & $\tau^2_U$ \\ 
  \hline
Posterior Mean & -0.316 & 0.027 & 222.294 & 16.332 \\ 
Posterior SD & 0.004 & 0.005 & 32.379 & 1.219 \\ 
   \hline
\end{tabular}
\caption{Posterior means and standard deviations of parameters in the post-hoc river model shown in equation \eqref{eq:post1}.}
\label{tab:post1}
\end{table}

\begin{table}[ht]
\centering
\begin{tabular}{rrrrrr}
  \hline
 & $\delta_0$ & $\delta_{Gauged}$ & $\delta_{Gauged:Managed}$ & $\tau^2_\beta$ & $\tau^2_U$ \\ 
  \hline
Posterior Mean  & -0.359 & 0.062 & 0.014 & 250.942 & 16.000 \\ 
Posterior SD & 0.006 & 0.009 & 0.006 & 40.548 & 1.157 \\ 
   \hline
\end{tabular}
\caption{Posterior means and standard deviations of parameters in the post-hoc gauged/managed model shown in equation \eqref{eq:post2}.}
\label{tab:post2}
\end{table}

Results from these two models allow us to answer research question 2.  Specifically, we conclude that: (1) the construction of the streamflow variable in \citet{erhardt2024homogenized} meaningfully captures hydrological information for both gauged and ungauged locations; (2) the relative importance of this hydrological variable on predicting drought is higher at locations presently ungauged and off major river networks; and (3) there is a low level of concern that water management practices confound these findings.  

\section{Discussion}
Our scientific goal was to quantify the relative statistical importance of a streamflow measurement among of a set of seven covariates on ordinal drought levels in the US, with a specific focus on controlling for autocorrelation, variability in variance, and classifications of different regions.  To answer our research questions, we fit a Bayesian hierarchical model with a latent Gaussian response variable that allowed for covariate effects, variability in variance, and temporal autocorrelation.  We selected a preferred seasonal variance structure and autocorrelation order.  After controlling for these, we confirmed that: streamflow was the most statistically important predictor on average; often the single largest magnitude predictor for most locations; and among the most statistically important predictors for most locations.  Through post-hoc analyses, we confirmed this effect was not an artifact of the spatial homogenization procedure used to derive \texttt{streamflow}, and that the statistical importance of this variable on drought is highest at ungauged locations off of major river networks.  

The streamflow variable came from USGS data and was therefore largely independent of the more commonly available gridded meteorological covariates drawn from NLDAS-2.  Accordingly, we might expect that the information contained in streamflow for modelling droughts notably extends information already avaiable in the suite of NLDAS-2 variables.  This is exactly what we see.  Based on our post-hoc analysis, streamflow gauges on smaller rivers and streams have slightly more importance than those on major rivers. This initial finding should be explored further, but they suggest that increased in-situ streamflow monitoring and more comprehensive streamflow models that estimate flows in ungauged rivers can increase the accuracy of empirical drought models. Future research could further test the spatial variability of $\beta_{streamflow}$ using additional ancillary datasets that represent hydrologic processes (e.g. surface topography, soil type, land use) and water management practices (e.g. presence of reservoirs, irrigation channels, interbasin transfers). Additionally, the drought model applied in this study could be tested with streamflow estimates from the National Water Model in place of $\beta_{streamflow}$ to assess the consistency of these results.

\section{Data and Code Availability}
All data are freely available at Data Dryad (\url{https://datadryad.org/stash/dataset/doi:10.5061/dryad.g1jwstqw7#citations}) with code available at \url{https://github.com/heplersa/USDMdata}.  The data are described in extensive detail in \citet{erhardt2024homogenized}.

\section{Funding Statement}
The authors acknowledge support from NSF award \#2151881.

\bibliographystyle{jasa}
\bibliography{droughtref} 

\newpage
\appendix
\counterwithin{figure}{section}
\section{Appendix/Supplemental}

\begin{figure}[!h]
  \centering
\includegraphics[width=6in]{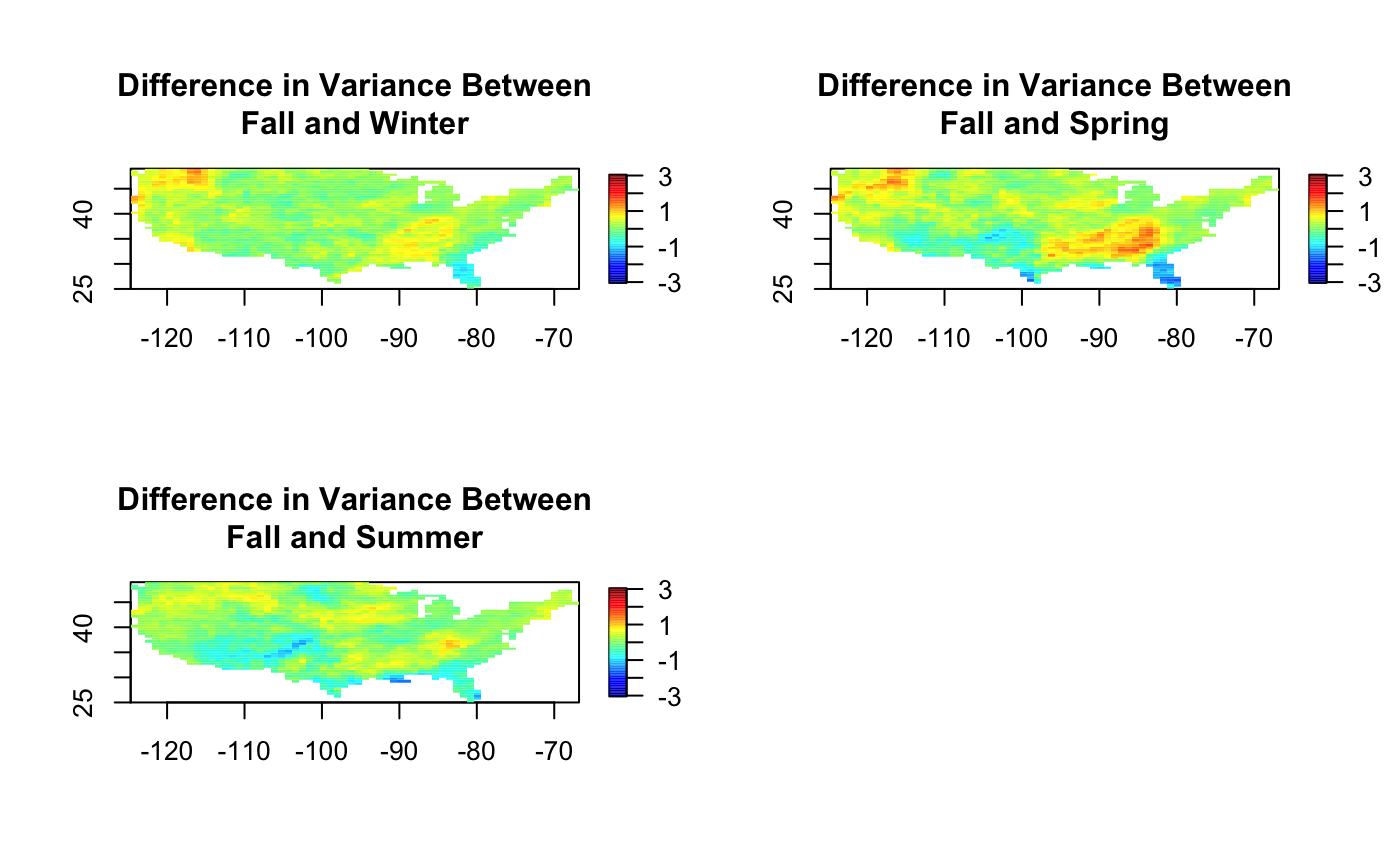}
    \caption{Difference between empirical variances of drought level in the fall and every other season.}
  \label{fig:fall}
\end{figure}

\begin{figure}[!h]
  \centering
\includegraphics[width=6in]{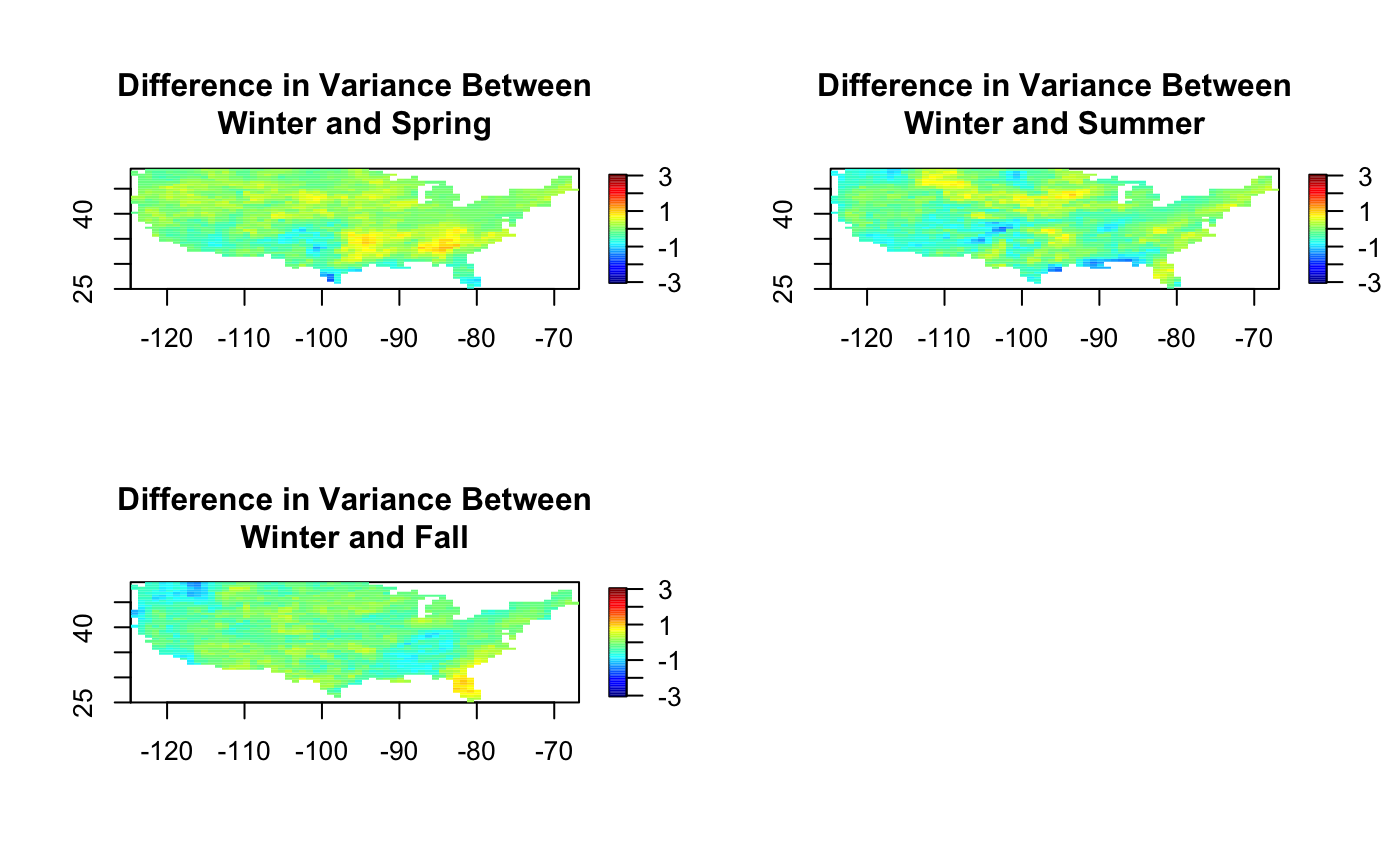}
    \caption{Difference between empirical variances of drought level in the winter and every other season.}
  \label{fig:wint}
\end{figure}

\begin{figure}[!h]
  \centering
\includegraphics[width=6in]{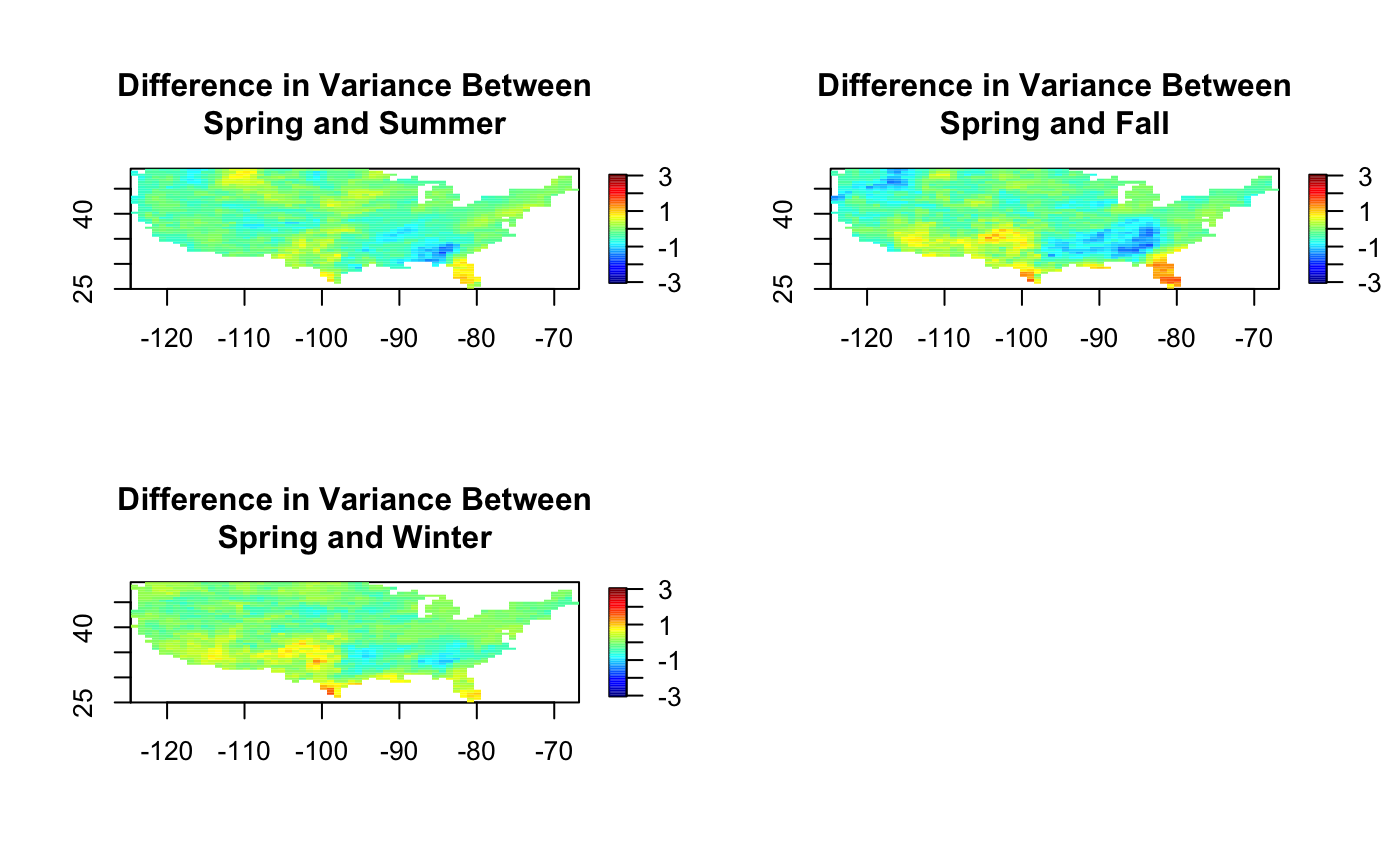}
    \caption{Difference between empirical variances of drought level in the spring and every other season.}
  \label{fig:spring}
\end{figure}

\begin{figure}[!h]
  \centering
\includegraphics[width=6in]{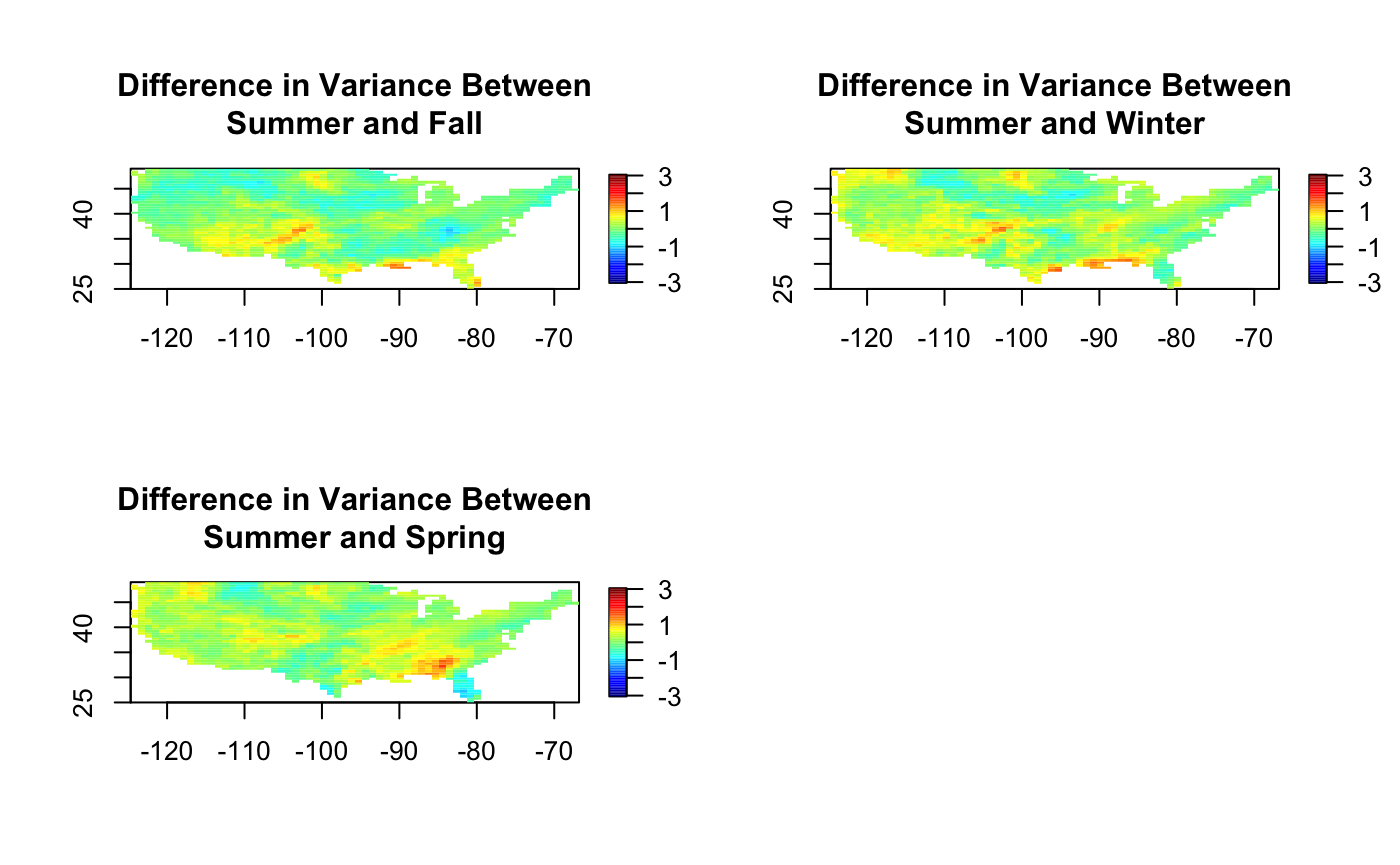}
    \caption{Difference between empirical variances of drought level in the summer and every other season.}
  \label{fig:summ}
\end{figure}

Figures \ref{fig:fall}, \ref{fig:wint}, \ref{fig:spring} and \ref{fig:summ} show differences in variances between all unique pairs of seasons, to help visualize changes to variance shown in Figure \ref{fig:vars}.

\begin{figure}[h!]
  \centering
    \includegraphics[width=\textwidth]{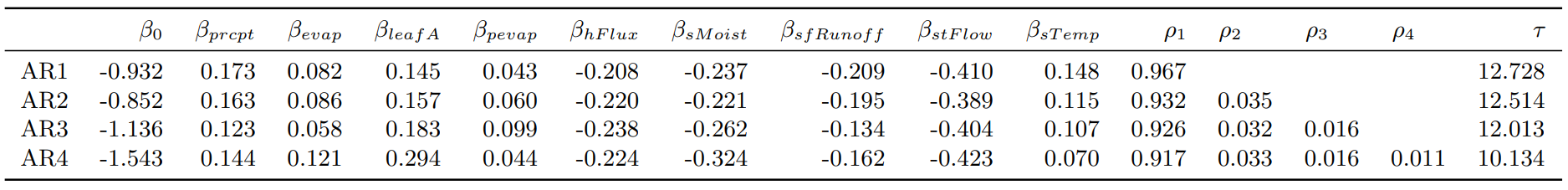}
    \includegraphics[width=\textwidth]{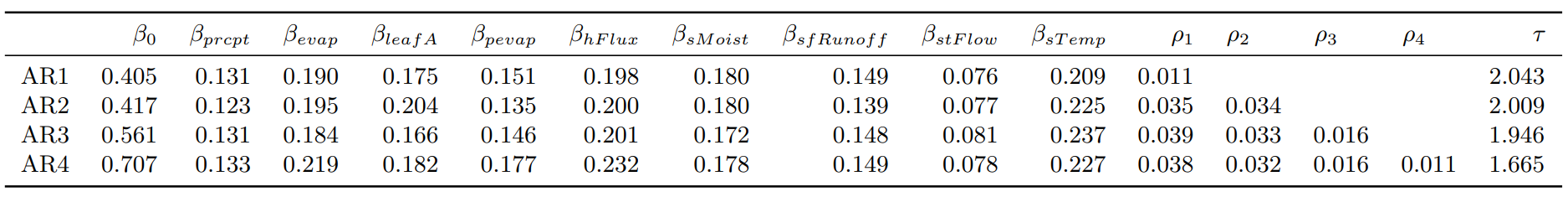} 
    \caption{Example posterior mean values (top) and posterior standard deviations (bottom) for the AR(k), k=1, 2, 3, 4 models when fit to grid cell $i=1$.}
    \label{fig:AR(k)}
\end{figure}

Figure \ref{fig:AR(k)} shows parameter estimates for a single grid cell's model fits for AR(1) through AR(4).  These preliminary models were fit to all locations, and used with WAIC to select an overall best model order which was AR(1).

\begin{figure}[h!]
  \centering
\includegraphics[width=5in]{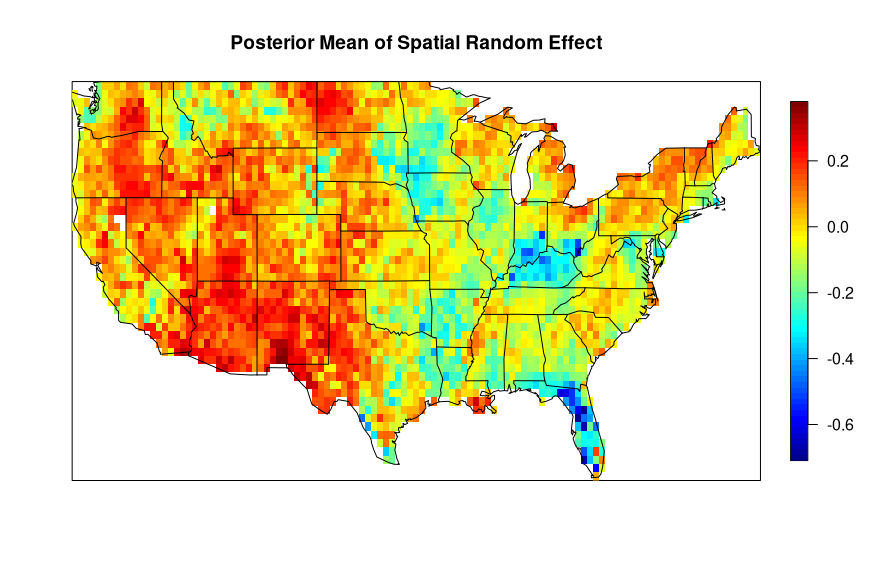}
    \caption{Posterior mean of the spatial random effect $\bfU$ in the model shown in equation \eqref{eq:post1}.}
  \label{fig:U}
\end{figure}

Figure \ref{fig:U} shows the posterior mean of the spatial random effect $\bfU$ from the model in equation \eqref{eq:post1}.  This effect captures unmeasured spatial variability in $\beta_{streamflow}$ across locations, and allows us to control for the spatial dependence on $\beta_{streamflow}$ when investigating the effects of the other indicators.  Thus, in this post-hoc step we capture some spatial dependence which we were unable to capture in the earlier model fit due to computational limitations.

\end{document}